\documentclass[10pt,aps,pra,twocolumn,superscriptaddress]{revtex4-1}

\usepackage{amssymb,amsmath}
\usepackage{graphicx}
\usepackage{color}
\usepackage{hyperref}
\usepackage{listings}
\usepackage{siunitx}
\usepackage{gensymb}
\usepackage{textcomp}
\usepackage{xspace}

\hypersetup{
    unicode=false,          
    pdftoolbar=true,        
    pdfmenubar=true,        
    pdffitwindow=false,     
    pdfstartview={FitH},    
    pdfauthor={none},     
    colorlinks=true,       
    linkcolor=blue,          
    citecolor=red,        
}
\graphicspath{{figures/}}

\newcommand{\micron}{{\textmu}m\xspace}

\begin{document}

\definecolor{dkgreen}{rgb}{0,0.6,0}
\definecolor{gray}{rgb}{0.5,0.5,0.5}
\definecolor{mauve}{rgb}{0.58,0,0.82}

\lstset{frame=tb,
  	language=Matlab,
  	aboveskip=3mm,
  	belowskip=3mm,
  	showstringspaces=false,
  	columns=flexible,
  	basicstyle={\small\ttfamily},
  	numbers=none,
  	numberstyle=\tiny\color{gray},
 	keywordstyle=\color{blue},
	commentstyle=\color{dkgreen},
  	stringstyle=\color{mauve},
  	breaklines=true,
  	breakatwhitespace=true
  	tabsize=3
}

\title{Electric-field noise from thermally-activated fluctuators in a surface ion trap}

\author{Crystal Noel}
\thanks{These authors contributed equally to this work.}
\affiliation{Department of  Physics, University of California, Berkeley, California 94720, USA}
\author{Maya Berlin-Udi}
\thanks{These authors contributed equally to this work.}
\affiliation{Department of  Physics, University of California, Berkeley, California 94720, USA}
\author{Clemens Matthiesen}
\thanks{These authors contributed equally to this work.}
\affiliation{Department of  Physics, University of California, Berkeley, California 94720, USA}
\author{Jessica Yu}
\affiliation{Department of  Physics, University of California, Berkeley, California 94720, USA}
\author{Yi Zhou}
\affiliation{Department of  Physics, University of California, Berkeley, California 94720, USA}
\author{Vincenzo Lordi}
\affiliation{
Lawrence Livermore National Laboratory, Livermore, CA, 94551, USA}
\author{Hartmut H\"affner}
\email{hhaeffner@berkeley.edu}
\affiliation{Department of  Physics, University of California, Berkeley, California 94720, USA}
\date{\today}

\begin{abstract}

We probe electric-field noise near the metal surface of an ion trap chip in a previously unexplored high-temperature regime. We observe a non-trivial temperature dependence with the noise amplitude at 1-MHz frequency saturating around 500~K. Measurements of the noise spectrum reveal a $1/f^{\alpha\approx1}$-dependence and a small decrease in $\alpha$ between low and high temperatures. This behavior can be explained by considering noise from a distribution of thermally-activated two-level fluctuators with activation energies between 0.35~eV and 0.65~eV. 
Processes in this energy range may be relevant to understanding electric-field noise in ion traps; for example defect motion in the solid state and surface adsorbate binding energies.
Studying these processes may aid in identifying the origin of excess electric-field noise in ion traps -- a major source of ion motional decoherence limiting the performance of surface traps as quantum devices.

\end{abstract}

\maketitle

\section{Introduction}
Electric-field noise is a major limiting factor in the performance of ion traps and other quantum devices \cite{Paladino20141/Information,Brownnutt2015}. Despite intensive research over the past decade, the nature and cause of electric-field noise near surfaces is not well understood. Progress in this field has wide-ranging applications for precision measurements close to surfaces. For instance, patch potentials affect Casimir force measurements \cite{Garrett2015TheMicroscopy,Xu2018ReducingMicrodevices} and limit the sensitivity of gravitational experiments that use test masses \cite{Antonucci2012InteractionMass,Armano2017Charge-InducedPathfinder}. Electric-field noise from surfaces is also detrimental to the performance of solid-state quantum bits in diamond \cite{Kim2015DecoherenceNoise,Myers2017Double-QuantumCenters}. For trapped ions, electric-field noise manifests as decoherence of the motional modes, which are generally relied on to entangle ions in the same trapping potential. It is one of the main obstacles to realizing a large-scale ion-trap quantum processor \cite{Eltony2016}.  A better understanding is not only important for the fabrication of low-noise quantum devices, but may also answer fundamental questions about the physics of noise at surfaces, such as the dynamics of defects and adsorbates. 

Experiments with ions in surface traps have explored the dependence of noise on frequency, distance to the surface, and temperature of the surface \cite{Turchette2000,Deslauriers2006a,Boldin2018,Sedlacek2018,Labaziewicz2008,Labaziewicz2008a,chiaverini2014a,Bruzewicz2015MeasurementTemperatures}. Research has also focused on the influence of trap material and surface treatments \cite{chiaverini2014a,Hite2012,Daniilidis2014,McKay2014}. The results of these studies have varied widely and often inconsistently between individual surface traps. Several models have been proposed to explain the electric-field noise and uncover physical noise mechanisms \cite{Turchette2000,Dubessy2009,Low2011,Daniilidis2011,Safavi-Naini2011,Safavi-Naini2013,Kim2017Electric-fieldExperiments,Kumph2016}, but the wide range of measurement results has made confirming or rejecting specific models challenging.

All previous measurements of the temperature scaling of surface trap electric-field noise were taken near and below room temperature \cite{Bruzewicz2015MeasurementTemperatures,Labaziewicz2008,chiaverini2014a,Sedlacek2018}. In many of these experiments, the temperature dependence follows a power-law scaling where the scaling exponent varies between 1.5 and 4. Below 70 K, some measurements show a plateau of the noise \cite{Bruzewicz2015MeasurementTemperatures,Labaziewicz2008a} while in another case the power-law exponent decreases \cite{chiaverini2014a}. In recent work, an Arrhenius scaling was observed after the trap was exposed to \textit{ex situ} ion milling \cite{Sedlacek2018}.

Here, we present the temperature dependence of electric-field noise in a surface trap significantly above room temperature (295 - 530 K).
We find that the electric-field noise saturates at high temperatures. This behavior can be explained by considering an ensemble of thermally activated fluctuators (TAFs) with a broad range of activation energies that peak around 0.5~eV. 
The TAF model, widely used to explain the temperature dependence of $1/f$ noise in metals and semiconductors \cite{Dutta1981a, Weissman19881fMatter,Paladino20141/Information}, ties together the temperature and frequency variables, requiring in our case that the saturation of noise be accompanied by a small change in the frequency dependence. We can experimentally confirm a small change in the $1/f$ scaling between room and high temperature, linking both frequency and temperature-scaling measurements of electric-field noise in our ion trap.

Applying the TAF model to our data, we can generate a distribution of fluctuator activation energies that produces a noise spectrum as a function of frequency and temperature consistent with our measurements. We can then consider processes in which these specific energy scales are relevant, and thus gain insight into which microscopic mechanisms might contribute to the electric-field noise in our system. Binding energies for adatom adsorbates, for instance, are in the relevant energy range, as are activation energies to motion of atomic defects in solid-state systems, which have been the subject of extensive research and are well described by the TAF model~\cite{Dutta1981a, Weissman19881fMatter,Paladino20141/Information}.

\begin{figure}[t!]
\centering
\includegraphics[]{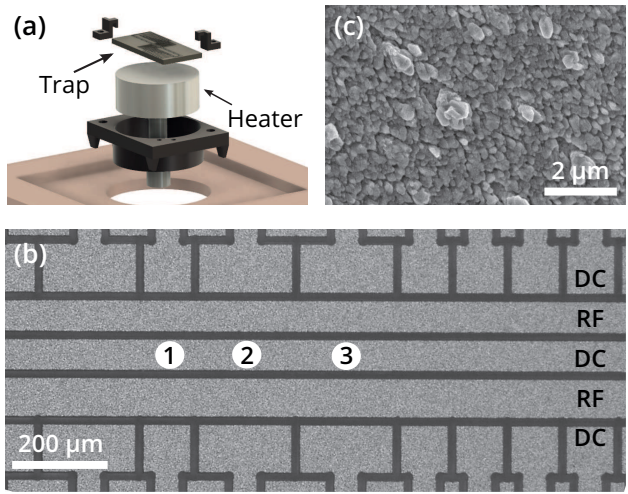}
\caption{(a)~Exploded rendering of the heater and trap set-up. The trap is clamped onto the button heater and wire-bonded to the chip carrier below. (b)~Grey-scale optical microscope image of the trap electrodes with numbers indicating the trapping locations. The electrodes are made of Al and Cu. (c)~Scanning electron microscope image of the surface of a similarly fabricated trap showing sub-micron structure. \label{fig:fig1}}
\end{figure}
  
\section{Experiment and results}
\label{sec:results}
In our experiment, we employ single $^{40}$Ca$^+$ ions in a linear surface trap as electric-field noise sensors. Relevant parts of the experimental setup, the trap geometry and structural information on the electrode metal surface are depicted in Fig.~\ref{fig:fig1}.
The trap chip is clamped onto a resistive heater (see Fig.~\ref{fig:fig1}(a) for an exploded view).
The ion trap is fabricated by laser-etching trenches of 100-\micron depth and 20-\micron width into a fused silica substrate to define the electrode geometry (Translume, Ann Arbor, Michigan) and evaporating layers of 15-nm titanium (Ti), 500-nm 
aluminum (Al), and 30-nm copper (Cu) on top. We repeat the metal evaporation process at two different angles, depositing 1.09~\micron of metal in total.

A grey-scale optical micrograph of the central area of the trap chip is shown in Fig.~\ref{fig:fig1}(b). Individual $^{40}$Ca$^+$ ions are trapped about 72~\micron above the surface along the central electrode. Here we report on measurements from three locations, as indicated by the labels. Prior to the measurements the trap chip underwent several cycles of exposure to atmospheric conditions and week-long bakes in the vacuum chamber at \ang{160} C. Figure~\ref{fig:fig1}(c) displays a scanning electron microscope (SEM) image of a similarly-fabricated trap, revealing a fine-grained structure.

Electric-field noise is measured by probing the ion motional heating rate. The spectral density of the electric-field noise $S$ is related to the heating rate $\dot{\bar{n}}$ of a motional mode by
\begin{equation}
S(\omega,T) = \frac{4m\hbar \omega}{q^2}\dot{\bar{n}}(\omega,T) ,
\label{eqn:hr}
\end{equation}
where $m$ and $q$ are the mass and the charge of the ion, respectively, $\omega$ is the motional mode frequency and $\hbar$ the reduced Planck constant. We measure heating rates of the axial mode which is parallel to the surface and the central electrode in Fig.~\ref{fig:fig1}(b) using the decay of carrier Rabi flops \cite{roosthesis}.

\begin{figure}[t!]
\includegraphics[]{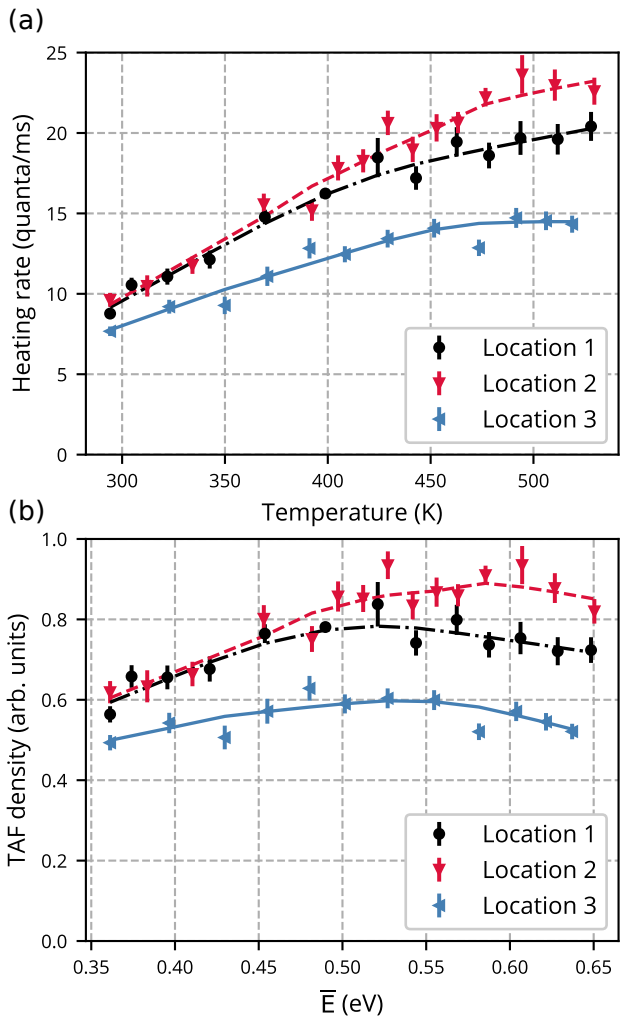}
\caption{(a) Heating rates of the axial mode at a frequency $\omega = 2\pi \times 1$~MHz as a function of trap temperature for the trapping locations labeled in Fig.~\ref{fig:fig1}(b). Data are shown as closed symbols, smoothing of the data gives the curves. (b) Density of thermally activated fluctuators with respect to their activation energy according to Eq. \eqref{eqn:Sfinal}. Data points are derived from the heating rate data in (a), the curves are from the smoothed data. Error bars represent one s.d. uncertainty.\label{fig:all3_temp}}
\end{figure}

First, we describe measurements of the temperature dependence of $S(\omega,T)$. The secular axial frequency is set to $\omega = 2\pi \times 1$~MHz here. The trap temperature is measured with a thermal imaging camera positioned outside the vacuum chamber, allowing us to determine the difference to room temperature with a systematic uncertainty of $\pm 10\%$ (see Appendix \ref{app:temp_uncertainty}).
We are able to measure heating rates from room temperature up to about 530 K. At higher temperatures ion lifetimes are less than a few minutes, which is too short to take measurements. Data from the three trapping locations are displayed in Fig.~\ref{fig:all3_temp}(a). The initial increase in heating rates between 300 and 400 K can be described by a power law $S(2\pi \times 1~ \mathrm{MHz}, T) \propto T^{1.8(1)}$, matching the dependence observed in some previous studies \cite{Labaziewicz2008a,Bruzewicz2015MeasurementTemperatures}. Above 400~K, the heating rates start to level off at all three trapping locations, overall roughly doubling between room temperature and 500 K.

We first fit the full range of our temperature scaling data to a power law for comparison to previous results. We find scaling exponents between 1.1 and 1.6 for the three data sets that are consistent with previous results \cite{Labaziewicz2008a,Bruzewicz2015MeasurementTemperatures}, but reduced $\chi^2$ values of 2.7-5.1 indicate that a power law is not a good fit (see Appendix \ref{sec:other_models} for more details). 

Some previous heating rate temperature dependencies have been found to be consistent with Arrhenius behavior \cite{Labaziewicz2008a,Sedlacek2018}  ($\dot{\bar{n}} = \dot{\bar{n}}_0 e^{-T_0/T}$). This behavior is predicted in two proposed models for electric-field noise in ion traps: diffusion of adatoms \cite{Kim2017Electric-fieldExperiments} or a specific regime of the adatom dipole model \cite{Safavi-Naini2011}. Arrhenius curves fit our data somewhat better than power laws (reduced $\chi^2$ values of 1.7 to 2.6). However, diffusion on a smooth, infinite, planar surface at high temperatures, as described in Ref. \cite{Brownnutt2015}, predicts a $1/f^2$ dependence, which is inconsistent with our data as presented in Sec. \ref{sec:freqscaling}.
The adatom model from Ref. \cite{Safavi-Naini2011} proposes noise due to phonon-driven fluctuations of bound adatoms and does find a parameter window where Arrhenius behavior and $1/f$ scaling concur. However, to describe our temperature scaling data in the context of this model, we must have phonon frequencies above the Debye frequency. In addition to being non-physical, these phonon frequencies correspond to a regime of the model where noise is independent of frequency at our measurement frequencies. This is inconsistent with our measured $1/f$ dependence (see Appendix \ref{sec:other_models} for more details).

The TAF model developed by Dutta, Dimon and Horn \cite{Dutta1979EnergyMetals} does not predict a specific functional form of the temperature scaling. It provides a mathematical description of noise from thermally activated two-level-fluctuators, linking both the temperature and the frequency dependence to the same distribution of activation energies. While the model has successfully captured the behavior in a range of solid-state systems \cite{Dutta1981a,Phillips1987, Weissman19881fMatter}, it has not been discussed much in the context of ion trap surface noise.

\section{TAF Model}
\label{sec:taf-model}
\begin{figure*}[t]
\centering
\includegraphics[scale = 1]{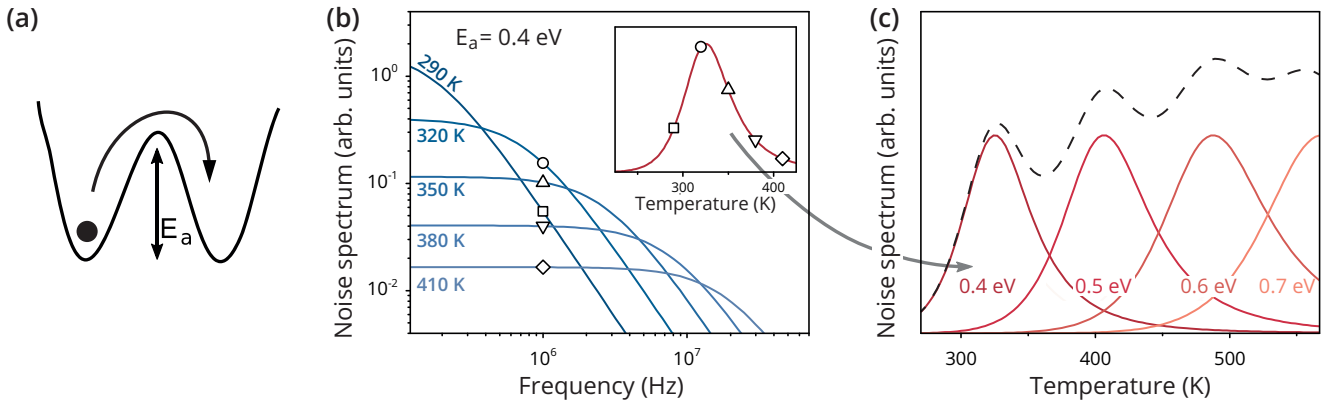}
\caption{Illustration of TAF spectral properties. (a) Cartoon picture of a TAF showing a charge hopping between two potential wells separated by an energy barrier $E_{\mathrm{a}}$. (b) Noise spectra for a single TAF with $E_{\mathrm{a}}=0.4$~eV at different temperatures. Inset:~Temperature dependence of noise at a frequency of $2\pi \times 1~\mathrm{MHz}$. Corresponding data points are displayed in the same shape. (c) Noise spectrum as a function of temperature for four TAFs (continuous curves) at $2\pi \times 1~\mathrm{MHz}$. The dashed line shows the sum.}
\label{fig:fig3}
\end{figure*}

A simple model for a TAF with a changing electric field is provided by a charged particle moving between two states separated by an energy barrier $E_\mathrm{{a}}$, see Fig.~\ref{fig:fig3} (a) \cite{Weissman19881fMatter}. Hopping between the sites gives rise to random telegraph electric-field noise. Assuming $E_\mathrm{a}\gg k_{B}T$, where $k_{B}$ is the Boltzmann constant, and, for simplicity, that the two sites are at the same energy, the characteristic transition rate for a single TAF is 
\begin{equation}
\Gamma = \frac{1}{\tau_0}e^{-E_\mathrm{{a}}/k_BT} ,
\label{eqn:gamma}
\end{equation}
where the hopping attempt time $\tau_0$ is of the order of an inverse phonon frequency, about $10^{-13}$ seconds \cite{Weissman19881fMatter}.
The resulting power spectrum is described by a Lorentzian centered at zero frequency, with a corner frequency $\Gamma$.

A trapped ion is sensitive to the component of the noise spectrum at its secular frequency $\omega$. Figure \ref{fig:fig3}(b) illustrates how the noise spectrum for a single TAF evolves with temperature, here using $E_{\mathrm{a}}=0.4$~eV for concreteness. When tracking the noise power at a fixed frequency $\omega$, it peaks at the temperature where $\Gamma = \omega$ and falls as the TAF fluctuates faster at higher temperatures (see inset).

Next, we consider the case where many TAFs with varying activation energies are present. Figure \ref{fig:fig3}(c) illustrates the noise power spectra for four independent TAFs, again at a fixed frequency of $2\pi \times 1~\mathrm{MHz}$ (continuous curves), and the resulting combined spectrum (dashed curve). At higher temperatures, fluctuators with higher energy barriers contribute to the noise at a fixed frequency.

For a continuous distribution of TAF energies $D(E_\mathrm{{a}})$ that varies slowly compared to $k_{B}T$ we can make the approximation \cite{Dutta1979EnergyMetals} 
\begin{equation}\label{eqn:Sfinal}
S(\omega,T) \propto \frac{k_B T}{\omega} D(\bar{E}), 
\end{equation}
directly linking the noise spectrum to the fluctuator distribution. Here, $\bar{E}$ represents the activation energy when $\Gamma = \omega$ for a given $T$.

Using Eq.~\eqref{eqn:Sfinal} we calculate $D(\bar{E})$ from the measured $S(\omega,T)$ \footnote{Equation \eqref{eqn:Sfinal} becomes less accurate at high frequencies and temperatures. We obtain a correction to $D(\bar{E})$ as detailed in \cite{Note1}, which is shown in Fig. \ref{fig:all3_temp}(b).}. The results are shown in Fig.~\ref{fig:all3_temp}(b). The temperature range of our measurements gives us access to fluctuators with activation energies in the 0.35-0.65~eV range. $D(\bar{E})$ evolves gently in this range, and shows a peak around 0.5-0.6~eV for the three locations. We note that the distribution function given by Eq. \eqref{eqn:Sfinal} is not unique, that is, other distribution functions can produce the same temperature dependence of $S(\omega,T)$. In particular, $D(E_{\mathrm{a}})$ may have sharper features than those we obtain in Fig.~\ref{fig:all3_temp}(b) (see Appendix \ref{app:model}).

\subsection{Frequency Scaling}
\label{sec:freqscaling}
Before discussing the distribution of activation energies in the context of physical processes, we examine the link between temperature and frequency scaling inherent to the TAF model:
Eq.~\eqref{eqn:Sfinal} gives a frequency dependence $S(\omega) \propto 1/\omega^{\alpha=1}$
only when $D(\bar{E})$ is constant. For a non-uniform distribution function, as we obtain in Fig.~\ref{fig:all3_temp}(b), the exponent $\alpha$ is both frequency- and temperature-dependent.
The frequency-scaling exponent is then given by \cite{Dutta1979EnergyMetals}:
\begin{equation}
\label{eqn:alpha}
\alpha (\omega,T) = 1-\frac{1}{\ln \omega \tau_0} \left(\frac{\partial \ln S}{\partial \ln T} -1\right) .
\end{equation}

\begin{figure}[t]
\centering
 \includegraphics[]{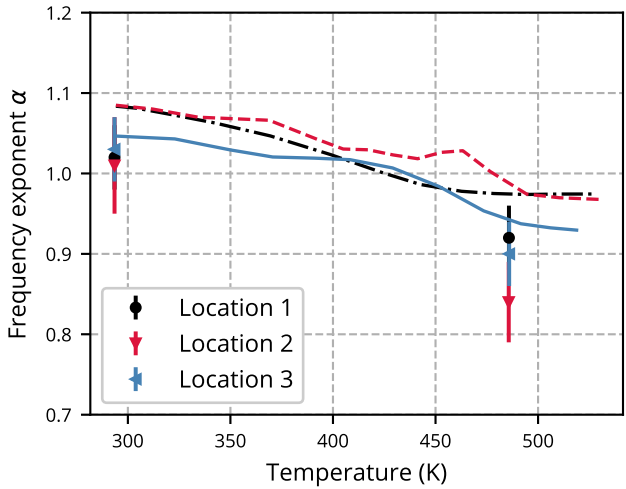}
\caption{Frequency exponent $\alpha$ as function of temperature. Data points represent frequency scaling at each trapping location. Curves give the predicted frequency scaling from Eq.~\eqref{eqn:alpha} and the smoothed temperature scaling data in Fig. \ref{fig:all3_temp}(a).}
\label{fig:alpha}
\end{figure}

If surface noise in our trap were accurately captured by two-level TAFs we should be able to deduce the electric-field noise frequency dependence from measurements of $S(T)$.
We compute $\alpha (2\pi\times 1~\mathrm{MHz},T)$ from the smoothed temperature-dependent data in Fig.~\ref{fig:all3_temp}(a) (solid curves) and display the results in Fig.~\ref{fig:alpha} (solid curves). Local regression smoothing with a span of about 0.6 allows us to 
estimate the change in $\alpha$ over the
temperature range of our measurements.
For all three locations $\alpha$ is predicted to be slightly above one at room temperature, and to drop to just below one at high temperatures. 

To test this prediction, we measure the noise frequency scaling at room temperature and at about 480~K. We confirm that the frequency dependence of heating rates in the range of about 0.6 - 1.2~MHz follows a power-law scaling by measuring across the full range (see \ref{fig:original_freqscaling} in Appendix \ref{app:freqscaling}). We then reduce the random error in the $\alpha$ estimate by measuring 10-20 heating rates for each trapping location at low and high temperatures, and low and high trap frequencies.

We plot the $\alpha$ estimates for the latter measurements as solid points in Fig.~\ref{fig:alpha}. All measurements are in the range typically associated with $1/f$ noise, but we also observe a statistically significant decrease in $\alpha$ across the three measurement locations when the trap temperature is raised. This trend is consistent with the predictions from Eq.~\eqref{eqn:alpha} suggesting that both our temperature and frequency-dependent data may be derived from the same fluctuator distribution.
We note a small offset between the prediction and the measurement which could be related to systematic uncertainties in our measurements  or be due to features in $S(T)$ which are not well resolved in our heating rate measurements (see Appendices \ref{app:methods} and \ref{app:model} for notes on systematic uncertainties and model uncertainties). We also note that the approximation Eq.~\eqref{eqn:Sfinal} relies on a smooth change in $D({E_{\mathrm{a}}})$ and is less accurate at the limits of our temperature range where the gradient is not well defined.

\section{Discussion and conclusions}
Relating data described by the TAF model to a noise mechanism relies on identifying physical processes with matching activation energies. For the energy range relevant in our case (0.35 - 0.65~eV) some adatom adsorption processes or diffusion energy barriers have activation energies in this range \cite{Dutta1981a,Ray2018}.

One physical mechanism in particular, defect motion in metal films, has been linked to $1/f$ noise in resistance measurements which is well described by the TAF model.
Metal film resistance noise is likely due to thermally activated defect migration around grain boundaries \cite{Dutta1981a,Weissman19881fMatter,Zhigalskii1997Films} and fluctuator energies in these systems peak in the 0.5 - 1~eV range, depending on the micro-structure and the metal in question \cite{Eberhard1978ExcessMetals,Dutta1981a}. Defects in aluminum are typically lower in energy than defects in other common metals, such as gold or silver \cite{Shewmon2016DiffusionSolids} and we find strong similarities between our results in Fig. \ref{fig:all3_temp} and corresponding data from resistance fluctuation measurements for polycrystalline aluminum films \cite{Koch19851fAlloys,Briggmann1994IrradiationinducedNoise}.

These similarities suggest a link between the microscopic mechanisms of resistance- and electric-field noise. However, there are clear differences in the way these two types of noise are measured:
In resistance noise measurements a current passing through the film is sensitive to defect motion in the bulk where conduction takes place, while electric-field noise in ion traps is dominated by surface processes \cite{Hite2012,Daniilidis2014,McKay2014}.
Defect motion is enhanced at grain boundaries, however, which includes the surface. For our trap, we expect not only a high defect density at the metal surface, but also charge traps in the oxide layer and adsorbate dipoles which can contribute to TAF noise. A trapped ion above the surface would be sensitive to their combined dynamics. 

In a wider context, TAF noise with different fluctuator distribution functions could account for some of the large variations of heating rates found in surface traps of different material and microstructure. However, additional noise sources may be present in other surface traps. For instance, the conceptually simple TAF model does not consider spatial information, or correlations between individual noise sources. Operation at a different temperature could also have a large impact on noise characteristics: At cryogenic temperatures, for example, quantum tunneling provides an additional mechanism for switching between the states of a two-level system~\cite{Phillips1987} and noise from these processes is known to cause decoherence in solid-state quantum systems, such as superconducting circuits~\cite{Paladino20141/Information}.

Considering defects at the surface as a noise source, it is interesting to think of treatments modifying their distribution function, thereby changing the electric-field-noise spectrum. For metals, annealing is a straightforward way to reduce defects~\cite{Pelz1985DependenceIrradiation,Briggmann1994IrradiationinducedNoise,Fleetwood1985DirectFilms} and some evidence has been reported already for the effectiveness of annealing trap electrodes~\cite{Labaziewicz2008}.
Ion milling can also strongly modify the surface defect density ~\cite{Pearton1990IonGaAs,Kiritani1994},
and we note that ion milling has been very successful at reducing noise in room-temperature traps \cite{Hite2012,Daniilidis2014}.

In conclusion, we have presented electric-field noise measurements in a surface ion trap in the temperature range of 295-530~K. Both the temperature and frequency-dependence of the noise are consistent with the same distribution of TAFs with activation energies around 0.5~eV.
While TAF noise can originate from many physical processes, we can gain insight into which processes are relevant to our system by considering those with energy barriers close to 0.5~eV, such as defect motion in the solid state.
The simple TAF model does not take spatial information or correlation between noise sources into account, but it provides a useful framework to classify average noise properties.
Understanding the influence of electrode materials and surface structure, such as metal grain size or oxide layers, surface roughness or the impact of surface treatments, such as ion milling and annealing, in the context of TAFs may further illuminate their microscopic origin, and inform the design of low-noise surface ion traps.

We thank Michael Crommie, Keith Ray, and John Clarke for insightful discussions and Matt Gilbert for taking the SEM image. Part of the trap fabrication was performed in the UC Berkeley Marvell Nanofabrication Laboratory. Part of this work was performed under the auspices of the U.S. Department of Energy by Lawrence Livermore National Laboratory under Contract DE-AC52-07NA27344. C.M. acknowledges partial funding from ONR via grant \#N000141712278 and M.B.-U. acknowledges an NSF Graduate Research Fellowship.

\appendix

\section{Experimental methods for temperature scaling data}
\label{app:methods}
The magnitude of electric-field noise at the ion location is measured through the heating of the ion's motion (see Eq. \eqref{eqn:hr}). The temperature of the motional mode $\bar{n}$ is obtained from measuring and fitting Rabi flops of the carrier transition 
following Doppler cooling and a variable wait time $t_{\mathrm{wait}}$ in the dark. A linear fit to $\bar{n}(t_{\mathrm{wait}})$ with typically 5-6 wait times yields the heating rate.

Data were taken at three locations on the trap chip approximately 200~\textmu m apart to find out how much the temperature dependence of electric-field noise varies across the surface trap.

The temperature scaling data presented in Fig. \ref{fig:all3_temp} were taken over the span of one week. For each trapping location in turn, the temperature was stepwise monotonically increased and heating rates were measured. Prior to taking the data in this manuscript, heating rates were measured across the same temperature range over the course of several weeks. Within the experimental uncertainty we observed no change in the temperature dependence of heating rates in subsequent measurements, indicating that raising the temperature of this trap to 550 K did not permanently alter its electric field noise characteristics.

\section{Details of frequency scaling exponent}
\label{app:freqscaling}

\begin{figure}[t]
\centering
\includegraphics[]{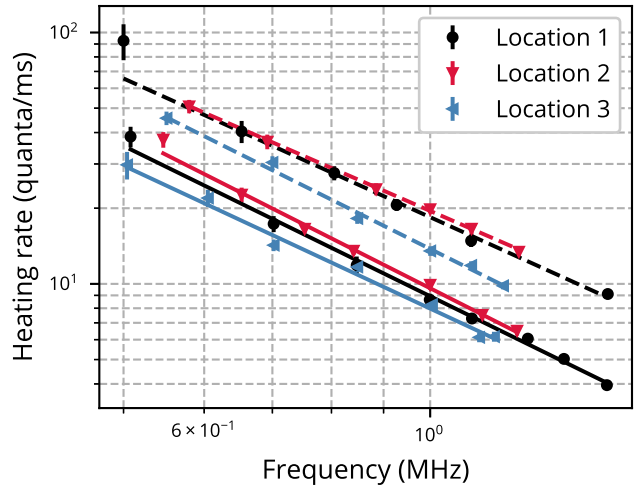}
\caption{Full range frequency scaling measurements at room temperature (solid lines) and high temperature (dashed lines) with a fit to $\dot{\bar{n}} \propto 1/\omega^{\alpha+1}$ for all three locations. At room temperature, $\alpha=0.99(7),1.05(8),0.89(9)$ for Locations 1-3, respectively. At elevated temperatures of 472~K, 485~K, and 480~K, $\alpha=0.83(9),0.76(9),1.02(8)$ for Locations 1-3, respectively.  No indications of technical noise such as peaks in the spectrum or deviation from a power law were ever observed.
\label{fig:original_freqscaling}}
\end{figure}

First, we present measurements of the noise spectrum over a wide range of frequencies for the three measurement locations at low and high temperatures in Fig.~\ref{fig:original_freqscaling}, where the trap frequency is varied between 500~kHz and 1.5~MHz. These measurements confirm that the frequency dependence follows a power-law scaling for all locations and both temperature limits.

Given this knowledge, we can then reduce the statistical uncertainty in the heating rate measurements for precise estimates of the scaling exponent $\alpha$ using a different measurement procedure: we cycle through the 5-6 wait times 10-20 times and fit all $\bar{n}(t_{\mathrm{wait}})$ to a single line. This procedure is performed for two frequencies. For each location and temperature, we chose the highest and lowest possible frequencies given the limitations of our electronics and the high heating rates at low frequencies. Taking data over extended periods of time reduces the influence of slow drifts in the experimental apparatus, such as the laser alignment, or drifts in the experimental parameters, such as laser power. The data for these measurements are presented in Fig.~\ref{fig:alpha}.

Systematic uncertainties do still arise for instance from fluctuations in the Rabi frequency: intensity noise on sub-second timescales on the laser driving the Rabi flops causes a faster decay of the Rabi flops which, for our experimental parameters, leads to a slight underestimate of heating rates at low trap frequencies and a slight overestimate of heating rates at high frequencies. Consequently, our calculation of the scaling exponent $\alpha$ likely underestimates the true value by 1-3~\%.

Finally, we quantify the change of the frequency scaling exponent $\alpha$ with temperature for all presented $\alpha$ measurements.

A t-test can be applied to the difference $\Delta \alpha$ for each pair of low and high temperature measurements. For an assumed true population mean of $\mu=0$ and a weighted standard deviation of the six samples (two at each location), the t-value is $t=3.95$. This t-value results in a $99.5\%$ confidence level rejection of the hypothesis that the frequency scaling remains unchanged at high temperatures when the location is varied. Instead, an average decrease of $0.12(3)$ is observed in the frequency scaling exponent. This results predicts that $\Delta \alpha$ would also be non-zero if we measured at other trapping locations.

\section{Note on technical noise}
\label{sec:tech}
Making a link between electric-field noise at the ion location and physical processes happening at the trap surface requires ruling out technical noise sources. Here we discuss several aspects of noise and trap characteristics which indicate our heating rate measurements are surface-noise limited. As seen in the frequency scaling measurements in Fig.~\ref{fig:original_freqscaling}, the heating rates decrease monotonically with frequency and are described well by a power-law fit. Technical noise is usually periodic in time and would show up as peaks in the noise spectrum which we have not observed.

Second, if we consider technical noise on the electrode voltages, we expect noise to be correlated over the size of individual (or multiple) electrodes. The magnitude of electric-field noise at the ion would then follow a distinct dependence on the position of the ion relative to the trap electrodes and the geometry of the electrodes. We have measured noise as a function of ion position along the trap chip and found no correlation between the amplitude of heating rates and the noise magnitude expected from electrode voltage noise.

Third, we consider Johnson noise, which is generated by all conductors in our setup. It is associated with a noise spectral density
\begin{equation}
    S_{\mathrm{V, JN}} = 4 k_{\mathrm{B}}TR(T)\; ,
\end{equation}
which has a white noise frequency spectrum, unlike the $1/f$ dependence we observe. We are therefore not dominated by Johnson noise, but we consider in the following what amount of noise it might contribute.
The Johnson noise sources closest to the ion are the trap electrodes and the heater underneath the trap chip, which are also the only elements heated significantly above room temperature.

If the trap electrodes are treated as a thin metal film, 
the resulting Johnson noise spectrum behaves as $S_{\mathrm{V, JN}} \propto~T^2$ \cite{DeVries1988,giancoli}, which is inconsistent with our temperature scaling measurements. A previous experiment was conducted in the same apparatus with a trap made of the same material, in which the electrodes were Ar$^+$ ion-milled \textit{in situ} to heating rates less than 0.01 quanta/ms \cite{Daniilidis2014}. 
This level of noise is three orders of magnitude lower than what we observe, and still exhibits $1/f$ behavior, suggesting an upper bound on the Johnson noise of that trap, and by extension the one used in this study.

For the heater, the resistance at a temperature of about 530~ K is 1.2~$\Omega$. The ion is trapped at a distance of 72~\textmu m from the trap surface and the trap chip is about 500-\textmu m thick, giving a distance of about 570~\textmu m between the ion and the heater surface. These parameters result in $S_{\mathrm{E, JN}} \approx 1.1\times10^{-13}\mathrm{V}^{2}\mathrm{m}^{-2}\mathrm{Hz}^{-1}$, which is three orders of magnitude lower than our measured electric-field noise.

\section{Effects of temperature uncertainty}
\label{app:temp_uncertainty}
\begin{figure}[t]
\centering
 \includegraphics[]{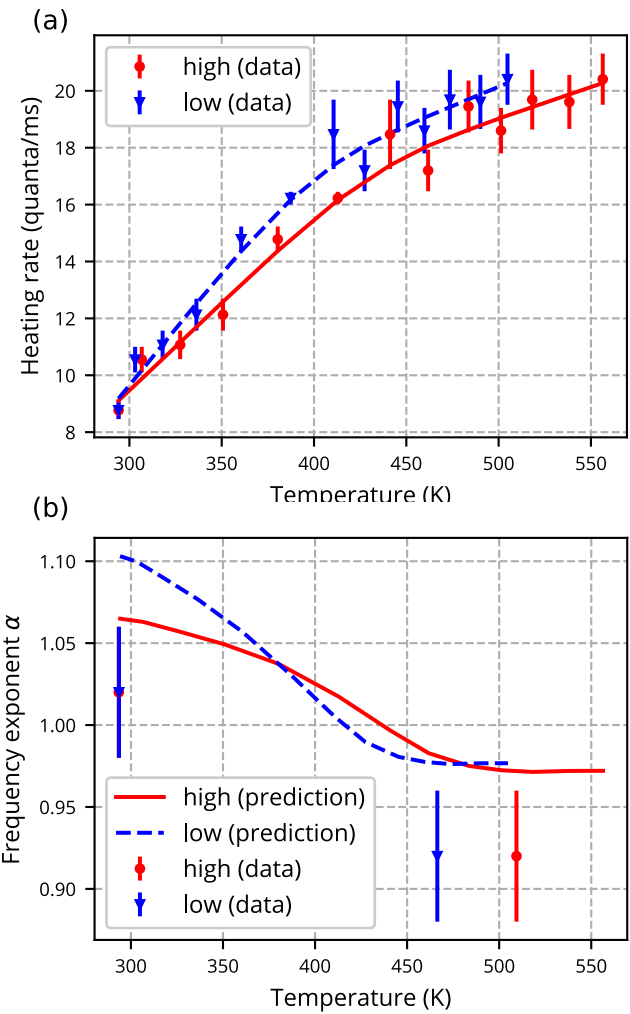}
\caption{Exploring the implications of uncertainty in trap temperature. We conservatively estimate a $10\%$ uncertainty in the emissivity of the reflection of the bottom glass of the trap based on independent calibrations. The `high' (`low') temperature data assumes a $10\%$ lower (higher) emissivity, and the data is shifted to the right (left). (a) The effect on the temperature scaling data of this uncertainty. The effect on the fluctuator distribution is qualitatively similar. (b) The effect of the uncertainty shifts the frequency exponent data in the same manner, but the prediction is different due to the new slope of the temperature scaling. \label{fig:uncertainty}}
\end{figure}

The trap temperature is estimated from thermal images of the trap/heater assembly taken with a thermal imaging camera (Seek Thermal XR) through an infrared viewport. The measured temperature is corrected for transmission losses through the infrared optics, which are independently calibrated. The most reliable temperature measurement is obtained from a reflection of the thermal emission from the trap glass substrate in the stainless steel mount below. We conservatively estimate the uncertainty in the emissivity of this glass reflection to be $\pm 10\%$ based on \textit{ex situ} calibrations, which gives the uncertainty in the difference to room temperature of $\pm 10\%$ quoted in Sec. \ref{sec:results}.
In Fig. \ref{fig:uncertainty} we explore the effect of the temperature uncertainty on our data and data interpretation. We consider the extreme cases of emissivity deviations of $+ 10\%$ (the low-temperature case) and $- 10\%$ (the high-temperature case) for data collected at location 1. The top panel of Fig. \ref{fig:uncertainty} replicates the temperature scaling from Fig. \ref{fig:all3_temp}: depending on the temperature calibration, the data are stretched to higher or compressed to lower temperatures. The effect on the fluctuator distribution is qualitatively similar, as we map temperature to fluctuator energy.

The frequency scaling exponent $\alpha$ is related to the slope of the noise spectrum (which is proportional to the heating rate), however, and $\alpha$ changes more subtly (see bottom panel of Fig. \ref{fig:uncertainty}): in the low-temperature case $\alpha$ is expected to change more strongly across the temperature range, while in the high-temperature case it varies more weakly.

\section{Details and constraints on the TAF model}
\label{app:model}
In this section we aim to give some more intuition about the thermally activated fluctuator model, the approximations introduced in Ref. \cite{Dutta1979EnergyMetals}, and limitations of the model.
In this model, a $1/f^{\alpha}$-noise spectrum is composed of the sum of spectra from individual, uncorrelated TAFs. A single TAF is associated with a Lorentzian noise spectrum, as described in Sec. \ref{sec:taf-model}, such that the frequency scaling exponent $\alpha$ varies smoothly and continuously from zero at low frequencies to two at high frequencies. Consequently, for an ensemble of fluctuators, $\alpha$ is also constrained to be within zero and two for the full spectral range. As the noise spectrum dictates the temperature dependence of noise at a fixed frequency, this condition also limits the shape of the temperature dependence that can be described by the TAF model.

For a more concrete picture, take Fig. \ref{fig:fig3}(c). Here, we considered four individual TAFs with activation energies of 0.4 to 0.7~eV. We see that a single TAF is associated with a certain width in temperature, and that the width is increasing with increasing temperatures (also with higher measurement frequencies). The curve for a single TAF determines the temperature resolution of the TAF model: any noise spectrum that has features sharper than these peaks, or a local rate of change higher than that of a single TAF cannot be described by the TAF model.

In the Dutta-Dimon-Horn (DDH) approximation of the TAF model, the link between the noise spectrum and a distribution of fluctuator energies is described by
\begin{equation}
    D_{\mathrm{DDH}} \propto \frac{S(\omega,T)}{T}.
    \label{eqn:Sfinal_app}
\end{equation}
In this approximation the amplitude of noise at a fixed frequency and temperature maps to the fluctuator density at a fixed energy. The approximation is most accurate in the limit of low temperatures and low frequencies, because a single TAF produces a very sharp peak in temperature in this limit. For the range of temperatures and frequencies explored in this work, the approximation works less well due to the large width of TAFs in temperature space, and there is no one unique fluctuator distribution that describes the noise temperature dependence. Eq. \eqref{eqn:Sfinal_app} then represents the smoothest fluctuator distribution.

\begin{figure*}[t]
\centering
\includegraphics[scale = 1]{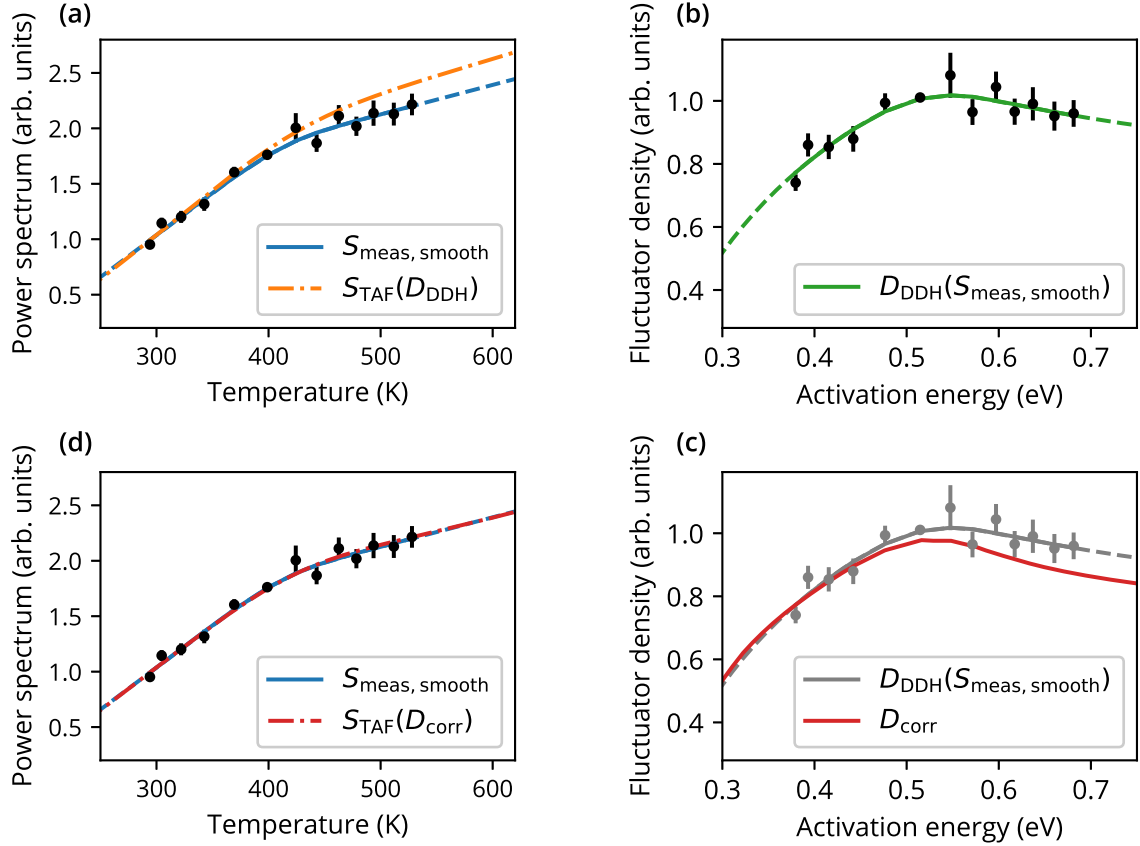}
\caption{Testing the Dutta-Dimon-Horn (DDH) approximation for the fluctuator distribution (Eq. \eqref{eqn:Sfinal_app}). (a) Heating rate data for location 1 (black points), and the smoothed curve (continuous blue curve, labelled $S_{\mathrm{meas,smooth}}$). The smoothed curve is extrapolated beyond our temperature range (dashed blue curve) based on the gradient at the low and high temperatures. (b) Using Eq. \eqref{eqn:Sfinal} we calculate the corresponding fluctuator distribution from the data (black points) and the smoothed curve (labelled $D_{\mathrm{DDH}}(S_{\mathrm{meas,smooth}})$). With $D_{\mathrm{DDH}}(S_{\mathrm{meas,smooth}})$ as input to the TAF model we evaluate the noise spectrum at 1-MHz frequency as a function of temperature and plot it in panel (a) (orange dash-dotted curve labelled $S_{\mathrm{TAF}}(D_{\mathrm{DDH}})$). For high temperatures the calculated spectrum overshoots the measured one by an increasing margin. We conclude the fluctuator distribution in (b) is not quite accurate. (c) Taking the ratio $S_{\mathrm{meas,smooth}}/S_{\mathrm{TAF}}(D_{\mathrm{DDH}})$ as a corrective factor for the distribution $D_{\mathrm{DDH}}$ we obtain a new estimate $D_{\mathrm{corr}}$ for the fluctuator density (red (lower) curve). (d) With $D_{\mathrm{corr}}$ as input to the TAF model we evaluate the noise spectrum again (red dash-dotted curve, labelled $S_{\mathrm{TAF}}(D_{\mathrm{corr}})$) and find good agreement with the data.
\label{fig:taf_correction}}
\end{figure*}

To check the validity of the distributions obtained by \eqref{eqn:Sfinal}, we calculate the noise spectrum in the TAF model $S_{\mathrm{TAF}}(D_{\mathrm{DDH}})$ from first principles using $D_{\mathrm{DDH}}$ as input and find deviations from the measured spectra that increase with temperature. For the highest temperatures in our measurements, $S_{\mathrm{TAF}}(D_{\mathrm{DDH}})$ overestimates the measured spectrum by about 10\%. For the data presented in Fig. \ref{fig:all3_temp}(b) we apply a first-order correction proportional to the ratio of the measured to calculated spectra. Using the corrected distribution $D_{\mathrm{corr}}$ of fluctuator energies as input to the TAF model $S_{\mathrm{TAF}}(D_{\mathrm{corr}})$, we then find agreement to better than 2\%. The procedure for obtaining a correction to the fluctuator distribution is illustrated in Fig. \ref{fig:taf_correction} for the data from location 1.

In summary, we note that, while the mapping from Eq. \eqref{eqn:Sfinal} can be applied to any noise spectrum,
\begin{enumerate}
    \item the approximation is increasingly inaccurate for high temperatures and frequencies, and
    \item not all noise spectra satisfy the constraints underlying the TAF model.
\end{enumerate}
For high frequencies and temperatures in particular, the noise spectrum and $\alpha$ should be calculated from the TAF model directly.

Finally, we will use the TAF model to fit data from location 3 and discuss the some of the uncertainties involved. 
We begin by choosing a functional form for the distribution of fluctuator activation energies. The function should be smooth and continuous, and, for the sake of fitting, be parameterized by only a few variables. Further, it should be flexible enough to describe a wide range of possible noise temperature dependences. We make the arbitrary choice here of parameterizing the distribution of fluctuators by a sum of equally spaced Gaussians with a fixed width. In the example discussed below we deliberately limit the number of Gaussians to five. Their center energies span the range of 0.3 and 0.65~eV, which is the range relevant to the experimental data. The Gaussian full-width at half-maximum is fixed to be equal to the separation of the fluctuator center energies (0.07~eV in this case). The amplitudes of the five Gaussian fluctuator distributions are then the only free parameters to fit the heating rate temperature dependence.
Using significantly more Gaussians, or a larger width leads to overfitting and the fitted amplitudes become meaningless. We note that a single Gaussian fluctuator distribution with amplitude, width and center energy as free parameters also provides a reasonable fit to our data. 

To get a sense of how much the fit changes when a different fluctuator basis is used, we shift all Gaussian distributions step-wise to higher energies and fit the data again. This shift, along with the arbitrary number of Gaussians, illustrates that a single distribution calculated from this model is not a unique description of the underlying physical distribution due to the nature of the broad peaks in the temperature spectrum caused by a single TAF.

\begin{figure*}[t]
\centering
\includegraphics[scale = 1]{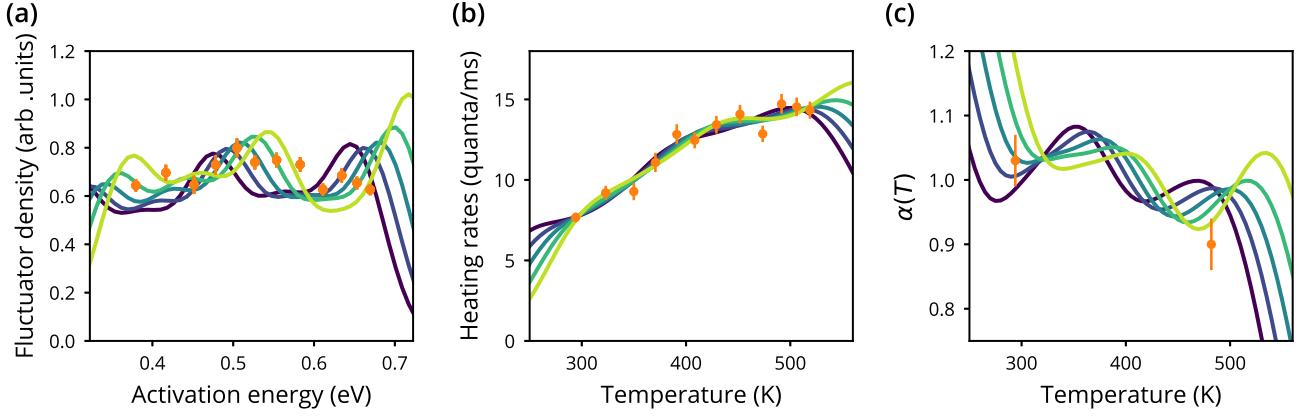}
\caption{First-principle calculations of the TAF model for the temperature scaling from location 3. Starting from a distribution of fluctuator energies (a), we calculate the noise spectrum (b), which is fit to the temperature scaling data, orange dots in (b), by varying the distribution of fluctuator energies. From the noise spectrum we also calculate the frequency scaling exponent, which is shown in (c). To ensure continuity of the distribution function, it is composed of the sum of equally spaced Gaussians and their amplitudes are the fitting parameters. The solid curves of different color are obtained by shifting the Gaussian center energies (see text for further details). The orange dots in (a) correspond to the approximation \eqref{eqn:Sfinal}. The data in (b) is fit well by all curves, despite their different underlying fluctuator distributions. The curves for $\alpha$ show some amount of spread, but share the downward trend in the 300-500 K range, which is also observed in the measurements (orange dots in (c) show the measurements for location 3). \label{fig:taf_fits}}
\end{figure*}

Figure~\ref{fig:taf_fits} shows the results of the fitting. Matching line colors link corresponding curves. In panel (a), we display the fluctuator distributions (solid curves) resulting from fitting the heating rate temperature dependence. The darkest curve corresponds to Gaussian center energies between 0.3 and 0.65 eV, while the lightest curve is for energies between 0.37 and 0.72 eV. Panel (b) shows the temperature dependence of the noise spectral density at 1-MHz frequency (solid curves), together with the measured heating rates (orange dots). Despite originating from different fluctuator distributions, the curves fit the data similarly well. Since we have no knowledge of the temperature dependence outside the measured region, we see the fitted curves fanning out at high and low temperatures.

Panel (c) displays the predicted temperature dependence of the frequency scaling exponent $\alpha$ from each curve in panel (a); the orange dots show the measured values for location 3. Since the prediction is derived from the gradient of the noise spectrum, the curves for $\alpha$ start to diverge beyond the temperature range of the data. Within the measured range, there is a clear downward trend for all curves, which we also see in our measurements.
In Fig. \ref{fig:alpha} text we presented a single prediction curve for $\alpha$ for the sake of clarity. However, being based on measurements, the $\alpha$ prediction from the TAF model is also associated with an uncertainty. The spread of the curves shown in Fig. \ref{fig:taf_fits}(c) illustrates the scale of this uncertainty, which would need to be taken into account for a quantitative comparison between the measurements and the $\alpha$ prediction. The downward trend in $\alpha$ predicted by the TAF model is a more robust feature, however, which matches our observations reasonably well.

\begin{figure}[ht]
\centering
\includegraphics[scale = 1]{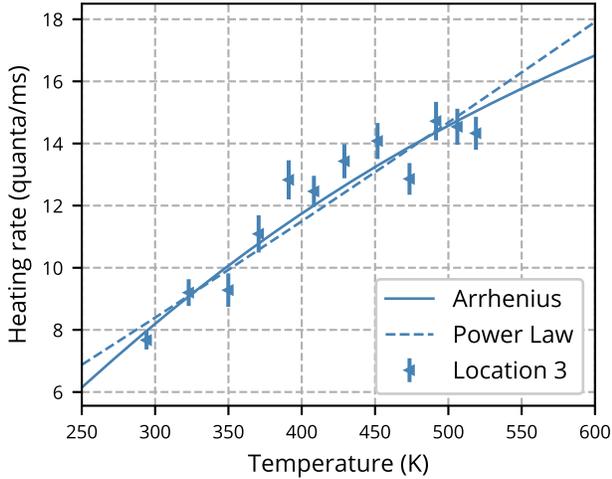}
\caption{Temperature scaling data from Fig. \ref{fig:all3_temp}(a) at location 3 with a fit to an Arrhenius curve (solid) and a power law (dashed). The resulting fit parameters are summarized in Table~I and the implications are discussed in `Other surface noise models'. In summary, neither curve describes the data well and the Arrhenius parameters observed are unlike other published results and not easily reconciled with the assumptions of the fluctuating dipole model. \label{fig:other_fits}}
\end{figure}

\section{Other surface noise models}
\label{sec:other_models}
Previous temperature scaling results have fit either a power law or Arrhenius-type behavior. We find that neither of these functional forms fit the data well, and even taking the fitted parameters at face value does not match any underlying physical model. The results of those fits are summarized in Table~I as well as Fig.~\ref{fig:other_fits}.

\begin{table*}[ht]
\centering
 \begin{tabular}{|c | c | c | c | c | c | c | c |} 
 \hline
 Location & \multicolumn{3}{|c|}{Power Law} & \multicolumn{4}{|c|}{Arrhenius}  \\
 &\multicolumn{3}{|c|}{$T^\gamma$} & \multicolumn{4}{|c|}{$e^{-T_0/T}$} \\
 \hline \hline
  & $\gamma$ & $\chi^2_P$ & $p_P$ & $T_0$& $E_b$ & $\chi^2_A$ & $p_A$\\ [0.5ex] 
 \hline
 1 & 1.4(1) & 5.1 & $5.7\times10^{-8}$ & 550(40) &0.047(3) &  2.6 & $2.8\times10^{-3}$\\ 
 \hline
 2 & 1.6(1) & 2.7 & $9.6\times10^{-4}$& 620(33) & 0.053(3) & 1.7 & $5.7\times10^{-2}$\\
 \hline
 3 & 1.1(1) & 2.8 & $2.1\times10^{-3}$& 430(40) & 0.037(3) & 1.9 & $3.8\times10^{-2}$ \\
 \hline
\end{tabular}
\caption{Summary of fit parameters for power law and Arrhenius dependence, including reduced $\chi^2$ values and $p$-values. \label{tab:fits}}
\end{table*}

A power law ($\dot{\bar{n}} = \dot{\bar{n}}_0 T^{\gamma}$) is a common result in heating rate temperature scalings. We find that the temperature exponent $\gamma$ is similar to those from \cite{Bruzewicz2015MeasurementTemperatures}, but does not provide a good fit to the data, as indicated by the values of the reduced $\chi^2$, $\chi^2_P$. Additionally the $p$-values are $10^{-3}$ or lower, which results in a $>99\%$ confidence rejection of this model. 
Arrhenius behavior ($\dot{\bar{n}} = \dot{\bar{n}}_0 e^{-T_0/T}$) can originate from diffusion of adatoms or a specific regime of the adatom dipole model. Previous published values of $T_0$ are in the range of 17-73~K \cite{Labaziewicz2008a}, 45~K and 63~K \cite{Sedlacek2018}. The values of $T_0$ for our data are an order of magnitude higher and we discuss the implications of those values in terms of the two models in the following.

Diffusion on a smooth, infinite, planar surface at high temperatures is described by \cite{Brownnutt2015} 
\begin{equation}
S_E \propto \frac{D_0 e^{-E_b/k_BT}}{\omega^2},
\end{equation}
where $E_b$ is the barrier to diffusion.
We measure a frequency scaling close to $1/f$, not a $1/f^2$ dependence predicted by this model. 

The adatom model proposed by \cite{Safavi-Naini2011}, posits a $1/f$ regime that would exhibit similar Arrhenius behavior. The physical mechanism behind the noise is phonon-driven fluctuations of bound adatoms. The frequency spacing of the vibrational states of the bound adatoms, $\nu_{10}$, depends on the adatom, but is typically around 1~THz. 
We can evaluate the model by fitting our data to  $\dot{\bar{n}} = \dot{\bar{n}}_0 e^{-T_0/T}$ and estimate in which frequency range this $1/f$ region would occur.

 The results of fits at each location are shown in Table~I, and $T_0$ is on average 530~K. This value $T_0$ would suggest that $\nu_{10} \approx 11$ THz , which is larger than the Debye frequency of 8 THz in aluminum \cite{Chipman1960}. In polycrystalline materials, such as our trap, the Debye temperature is even lower due to the excess volume \cite{Wagner1992}, ruling out phonon-driven noise mechanisms.

We can also use the average of value of $T_0$ to estimate where the $1/f$ region of the noise spectrum would occur in terms of frequency. The expression from Ref. \cite{Safavi-Naini2011} for the transition frequency of the lowest two bound vibrational states (the zero temperature transition), $\Gamma_0$, depends on properties of the material, and the bound adatom :
\begin{equation}
\Gamma_0 = \frac{1}{4\pi} \frac{\nu_{10}^4 m}{v^3 \rho} ,
\end{equation}
and only holds if $\nu_{10} < \omega_D$, the Debye frequency of the bulk material. In aluminum, the speed of sound is $v = 6320$ m/s, and the density is $\rho = 2.7$ g/cm$^3$. For these parameters, $\Gamma_0 = 10$ THz. The $1/f$ region begins around $\omega_c \approx \Gamma_0[1+(e^{h \nu_{10}/k_B T}-1)^{-1}]$, which for the highest temperatures measured in this work, is approximately $1.6 \times \Gamma_0 = 16$~THz, suggesting that in our measurement range the frequency scaling would still be flat. 

\subsection*{Dipole extension of TAF}
Here we discuss the number of fluctuating dipole noise sources needed to reproduce the electric-field noise level in our trap.
Electric field fluctuations parallel to the surface due to an averaged dipole fluctuation spectrum $S_{\mu}$ are given by \cite{Brownnutt2015}
\begin{equation}
S_E = \frac{3 \pi}{4 \sigma_d} \frac{1}{(4\pi \epsilon_0 d^2)^2} S_{\mu},
\end{equation}
where $\sigma_d$ is the areal density of dipoles, and $d$ is the distance to the electrode surface.
For a distribution of energy barriers $D(E_b)$,
\begin{equation}
S_{\mu} = \mu^2 \frac{\pi k_B T}{4 \omega} D(E_b = -k_B T \log (\omega \tau_0)).
\end{equation}
Assuming a Gaussian form for $D(E_b)$ and $\mu=5$~D, then $\sigma_d \approx 7-10 \times 10^{18}$ m$^{-2}$, or approximately 7-10 TAF dipoles per square nanometer. We note here that the roughness of the trap surface (see Fig. \ref{fig:fig1}) increases the effective surface area and that noise sources near the surface may play a role, both making a volumetric density a maybe more appropriate measure.

\bibliography{references}

\end{document}